\documentclass[twoside,fleqn]{article}
\usepackage{epsf}
\usepackage{rotate}
\usepackage{espcrc2}

\newcommand{\AmS}{{\protect\the\textfont2
  A\kern-.1667em\lower.5ex\hbox{M}\kern-.125emS}}
\newcommand{\be}[1]{\begin{equation} \label{(#1)}}
\newcommand{\ee}{\end{equation}}
\newcommand{\ba}[1]{\begin{eqnarray} \label{(#1)}}
\newcommand{\ea}{\end{eqnarray}}
\newcommand{\nn}{\nonumber}

\title{ Anomaly-Induced Effective Action and Inflation }

\author{
Julio C. Fabris 
\address{
Universidade Federal de Espir\'{\i}to Santo, Brazil
\\fabris@cce.ufes.br}, 
Ana M. Pelinson 
\address{ Centro Brasileiro de Pesquisas F\'{\i}sicas, Rio de Janeiro, 
Brazil
\\ana@fisica.ufjf.br} 
and 
Ilya L. Shapiro \thanks{Authors are grateful to CNPq (Brazil) for 
permanent support.}
\address{
Universidade Federal de Juiz de Fora, Brazil
\\ 
Tomsk Pedagogical University, Russia
\\shapiro@fisica.ufjf.br}}
       
\begin{document}

\begin{abstract}
In the early Universe matter can be described as 
a conformal invariant ultra-relativistic perfect fluid, 
which does not contribute, on classical level, to the 
evolution of the isotropic and homogeneous metric. If 
we suppose that there is some desert in the particle 
spectrum just below the Planck mass, then the effect 
of conformal trace anomaly is dominating at the 
corresponding energies. With some additional constraints 
on the particle content of the underlying gauge model 
(which favor extended or supersymmetric versions of the 
Standard Model rather than the minimal one), one arrives 
at the stable inflation. We review the model and report 
about the calculation of the gravitational waves on the 
background of the anomaly-induced inflation. The result 
for the perturbation spectrum is close to the one for 
the conventional inflaton model, and is in agreement 
with the existing Cobe data.
\end{abstract}

\maketitle

\section{Introduction}

Inflation is a necessary component of the cosmological 
standard model. Besides to present a simple 
solution to the flatness and horizon problems, such issues as 
metric and density perturbations  have been 
successfully studied in the context of the inflationary model
and led to a consistent scenario for
structure formation as well as the anisotropies in the 
relic radiation.
But, the inflaton potentials which are 
necessary to achieve the successful inflation 
are phenomenological potentials, which can be hardly derived 
from some quantum field theory. Therefore, there is a lack of 
a natural model for inflation. An alternative approach to inflation
can be based on the trace anomaly and on the 
effective action of gravity resulting from the 
quantum effects of matter fields on the classical gravitational 
background \cite{fhh,star,vile,anju,wave,hawk}.

\section{Inflationary solution}

Let us describe the general physical input \cite{star,vile,anju}.
Suppose there is a desert in the spectrum of 
particles which extends to some orders of magnitude 
below the Planck scale.
In the very early Universe the matter 
may be described by the free radiation, that is, microscopically, 
by the set of massless fields with negligible interactions between 
them. Due to the conformal invariance, 
these fields decouple from the conformal factor of the
metric. In this situation the dominating quantum 
effect is the trace anomaly which comes from the renormalization 
of the conformal invariant part of the vacuum action
(see \cite{birdav,book} for the introduction) 
\begin{equation}
S_{vacuum} = 
\int d^4x\sqrt{-g}\{ a_1 C^2 + a_2 E + a_3 {\nabla}^2 R\}. 
\end{equation}

The 
anomaly-induced effective action can be found explicitly 
\cite{rei,frts,book} with accuracy to 
an arbitrary conformal functional which vanishes for the special
case of the conformally flat metric.
The expression for the anomaly is:
\ba{anomaly}\nn
 <T_\mu^\mu> &=& 
- {{2} \over {\sqrt{-g}}}
\,g_{\mu\nu} {{\delta  {\bar \Gamma}}\over {\delta g_{\mu\nu}}}= 
\\&=&  
- {1 \over {(4\pi)^2}}\,(wC^2 - bE + c{\nabla}^2 R),
\ea
where $C^2, E$ are the square of the Weyl tensor and the 
integrand of the Gauss-Bonnet term, and
\begin{equation}
w=\frac{N_0}{120} + \frac{N_{1/2}}{20}+ \frac{N_1}{10} 
\end{equation}
\begin{equation}
b=\frac{N_0}{360} + \frac{11 N_{1/2}}{360} + \frac{31 N_1}{180}
\end{equation}
\begin{equation}
c=\frac{N_0}{180} + \frac{N_{1/2}}{30} - \frac{N_1}{10}\,,
\end{equation}
with $N_{0,1/2,1}$ - number of fields of spin $(0,1/2,1)$.
The solution for the effective action can be
written in terms of new variables 
$g_{\mu\nu} = {\bar g}_{\mu\nu}\cdot e^{2\sigma}$ as:
\ba{EA}\nn
{\bar \Gamma} &=& S_c[{\bar g}_{\mu\nu}] 
+\frac{1}{(4\pi)^2}\,\int d^4 x\sqrt{-{\bar g}}\,\Big\{ 
a\sigma {\bar C}^2 - 
\\&-&\nn
b\sigma ({\bar E}
-\frac23 {\bar {\nabla}}^2 {\bar R})
- 2b\sigma{\bar \Delta}_4\sigma -
\\&-&
\frac{3c-2b}{36}\,[{\bar R} 
- 6({\bar \nabla}\sigma)^2 
- 6({\bar \nabla}^2 \sigma)]^2)\Big\},
\ea
where
$$
\Delta_4 = {\nabla}^4 + 2\,R^{\mu\nu}\nabla_\mu\nabla_\nu 
- \frac23\,R{\nabla}^2 + \frac13\,(\nabla^\mu R)\nabla_\mu
$$
is the fourth derivative, conformal invariant
and self-adjoint operator and
$S_c[{\bar g}_{\mu\nu}]$ is some unknown functional of 
the metric  ${\bar g}_{\mu\nu}(x)$ which serves as an 
integration constant for the effective action.
This action includes some
arbitrariness, which was extensively investigated recently. 
It is accepted that all the arbitrariness is inside
the conformal functional $S_c[{\bar g}_{\mu\nu}]$.

The induced action can be presented in 
a nonlocal but covariant form using the original metric 
and then in a local covariant form via 
auxiliary scalars \cite{balbi}.
The most useful local form of the action is:
\ba{localEA}\nn
{\bar \Gamma} &=& S_c[g_{\mu\nu}] - 
\frac{3c-2b}{36(4\pi)^2}\,
\int d^4 x \sqrt{-g (x)}\,R^2 +
\\&+& \nn
\int d^4 x \sqrt{-g (x)}\,\left\{\,\,
\frac12 \,\varphi\,\Delta_4\,\varphi 
- \frac12 \,\psi\,\Delta_4\,\psi
\right.
\\&+& \nn
\left.
\varphi\,\left[\,\frac{\sqrt{b}}{8\pi}\,(E -\frac23\,{\nabla}^2R)\,
- \frac{w}{8\pi\sqrt{b}}\,C^2\,\right] 
\right.
\\&+& 
\left. 
\frac{w}{8\pi \sqrt{b}}\,\psi\,C^2 \,\right\}\,,
\ea
where $\varphi,\psi$ are auxiliary fields 
\cite{rei,balbi,wave}.

Consider the inflationary solution for the theory 
\begin{equation}
S_{total}\, =\, -\, M^2_P\,\int d^4x\sqrt{-g}\,R 
+ {\bar \Gamma}\,,
\end{equation}
where $M^2_P = {1}/{16\pi G}$ is the square of the 
Planck mass, and 
the quantum correction $\,{\bar \Gamma}\,$ is taken 
from the anomaly-induced action.
Since such a strategy was justified in the black hole 
case \cite{balbi}, one can
set $S_c[g_{\mu\nu}]=0$, including to it, when it is
not indicated explicitly, also the classical vacuum term. 

We look for the isotropic and homogeneous 
solution $g_{\mu\nu} = a^2(\eta)\,{\bar g}_{\mu\nu}$, where 
$\,\eta\,$ is conformal time. It proves useful to denote
$\,\sigma = \ln a$. We shall consider the conformally 
flat background and thus set 
$\,{\bar g}_{\mu\nu} = \eta_{\mu\nu}$.

The equations for the auxiliary fields:
\begin{equation}
\Delta_4\,\varphi 
+ \frac{\sqrt{b}}{8\pi}\,(E -\frac23\,{\nabla}^2R)\,
- \frac{w}{8\pi\sqrt{b}}\,C^2 = 0
\end{equation}\begin{equation}
\Delta_4\,\psi - \frac{w}{8\pi\sqrt{b}}\,C^2 = 0\,.
\end{equation}
reduce to 
\begin{equation}
{\nabla}^4\,\varphi 
+ \frac{\sqrt{b}}{2\pi}\,{\nabla}^4\sigma = 0\,,
\,\,\,\,\,\,\,\,\,\,\,\,\,\,\,\,\,\,\,\,\,
{\nabla}^4\,\psi = 0\,.
\end{equation}
The solutions 
\begin{equation}
\varphi = - \frac{\sqrt{b}}{2\pi}\,\sigma + \varphi_0\,,
\,\,\,\,\,\,\,\,\,\,\,\,\,\,\,\,\,\,\,\,\,
\psi =  \psi_0\,.
\end{equation}
where $\varphi_0,\,\psi_0$ are general solutions of the 
homogeneous equations
\begin{equation}
{\nabla}^4\,\varphi_0=0\,,\,\,\,{\nabla}^4\,\psi_0=0\,.
\end{equation}
Thus one meets an arbitrariness 
related to the choice of the 
initial conditions for the auxiliary fields.
Indeed, the inflationary solution does not depend nor on 
$\varphi_0,\,\psi_0$ neither on $S_c[g_{\mu\nu}]$.
The difference shows up only when we investigate the
metric perturbations. 

In terms of the 
physical time $t$, defined, as usual, through 
$a(\eta)d\eta = dt$ and
$H(t)= {\dot a}(t)/a(t) = {\dot \sigma}(t)$,
the equation is
\ba{URAVNEN}\nn
{\stackrel{...} {H}} &+&
7 {\stackrel{..} {H}}H
+ 4\,\left(1 + \frac{3b}{c}\right)\,
{\stackrel{.} {H}}H^2 + 4\,{{\stackrel{.} {H}}}^2 
\\ &+& 
\frac{4b}{c}\,H^4 - \frac{2M^2_{P}}{c}\,
\left(\,H^2 + {\stackrel{.} {H}}\,\right)  = 0\,.
\ea
The special solution corresponding to constant $H$:
\begin{equation}
H = \pm \frac{M_P}{\sqrt{b}}\,,\,\,\,\,\,\,\,\,\,\,\,\,\,\,
a(t) = a_0\cdot \exp {Ht}\,.
\end{equation}
Positive sign corresponds to inflation.

This solution was discovered and investigated by Starobinsky
and others in 80'th \cite{star,vile}. Also, there are
two other similar solutions for the FRW metric with 
$\,k=\pm 1\,$.
The special inflationary solution 
is stable with respect to the variations
(not necessary small) of the initial data for $a(t)$, 
if the parameters of the underlying quantum theory 
satisfy the condition $\frac{b}{c} > 0$, that leads to
\begin{equation}
N_1\, <\, \frac13\,N_{1/2}\, + \,\frac{1}{18}\, N_0\,.
\end{equation}
This constraint is not satisfied
for the MSM with
$\,N_1=12,$ $\,N_{1/2}=24\,$ and $\,N_0=4$. 

However, one can consider some aspects of the neutrino 
oscillations as an indication that the MSM should 
be extended. Below we consider two versions,
each of them leads to stable inflation.  

i) Extended SM: $N_1=12,\,N_{1/2}=48,\,N_0=8$

and

ii) MSSM $N_1=12,\,N_{1/2}=32,\,N_0=104$.

The advantage of stable
inflation is that it occurs independent of the initial data.
After the Big Bang, when the Universe started to expand and the 
typical energy decreased below the Planck order,
we can imagine some kind of "string phase transition".  
Starting from this point, the effective quantum field
theory is an adequate description, and the anomaly-induced
model applies. In case of the stable inflation,
the initial data for $a(t)$ and its derivatives 
do not need to be fine tuned, if only the condition 
$18N_1 < 6N_{1/2}+ N_0$ 
is satisfied -- the inflation is unavoidable. 

Let us calculate the necessary duration of 
inflation. Suppose we want the Universe to expand in $n$
e-folds. Then the total rate of inflation, in the 
Planck units, is 
\begin{equation}
\frac{a(t_0 + \delta t)}{a(t_0)} =
\exp\,\left\{\,4\pi\,
\sqrt{\frac{360}{N_t}}\,\,\delta t\right\} 
\end{equation}
where $\,N_t=N_0+11\cdot N_{1/2}+62\cdot N_1\,,$ and thus 
\begin{equation}
\,\delta \,t = \frac{1}{4\pi}\,\sqrt{ \frac{N_t}{360}}\cdot n\,.
\end{equation}
The time necessary
for 65 e-folds is around ten Planck times only. 

The numerical study has shown that the exponential solution
stabilizes in much shorter time. One can safely
derive the metric perturbations on the exponentially 
inflating background, independent on the initial data.

The most difficult question is how the inflation 
ends. So far, we do not have a definite answer to this 
question, but there are some particular indications that 
a solution is possible if we take the masses of the
matter particles (perfect fluid) into account \cite{fhh}.
A very important  
observation is that even when the Universe expands so rapidly, 
the average energy of the particles decreases. 
At some instant it  decreases such that their masses 
become relevant and then the matter part of the equation 
gets some dust component. In this 
case inflation is not anymore a solution. The classical 
solution for dust $a(t) \sim t^{2/3}$ also is not a solution 
because of the quantum term.
However, in this case both 
Einstein and matter terms behave like $t^{-2}$ while the
"quantum" part behaves like $t^{-4}$, 
and very rapidly the
anomaly-induced quantum term becomes irrelevant. 
Thus, one
can suppose that at the long-time limit $a(t) \sim t^{2/3}$ 
is a good approximation for the unknown solution of the 
equation with matter. Of course, the above consideration is not 
a solution of the problem, but  
one can hope that the solution will be found along this line.

Let us now review the equation for the metric perturbations
which is based on the bilinear expansion of the action
\cite{anju}.

Now we have to fix the arbitrariness 
related to the homogeneous 
solutions $\varphi_0$ and $\psi_0$ in the inflationary solution. 
The choice of initial data for $\varphi$ and $\psi$ defines the 
vacuum state for the perturbations. But, what could be
the criterion for the choice?

One can make a very useful comparison
with the vacuum of the black hole background \cite{balbi}. 
In the black hole case (semi-classical approach) the
vacuum
which provides a smooth transition to the Minkowski vacuum
at the space infinity is the Boulware one. 
Let us suppose that 
the proper cosmological vacuum for the expanding Universe 
reduces to the Minkowski one at infinite time.
Then the detailed consideration \cite{wave} gives 
\begin{equation}
\varphi = - \frac{\sqrt{b}}{2\pi}\,\sigma \,,
\,\,\,\,\,\,\,\,\,\,\,\,\,\,\,\,\,\,\,\,\,
\psi =  0\,.
\end{equation}
With this choice, one can perform the numerical study 
of the equation for the perturbations. The technical details 
can be found in \cite{wave}.

A numerical analysis requires that 
the equations must be dimensionless and the initial 
conditions must be consistently chosen. For the first point,
if one sets the Planck mass equal to one, time is 
automatically measured in the Planck units.  
For the initial conditions, we consider that the 
perturbations have a quantum origin: the seed of the 
perturbations are 
quantum fluctuations of the primordial fields \cite{wave}. 

Using the dimensionless equation and taking the initial conditions
consistent with
an initial quantum spectrum, one can 
integrate the fourth order equation numerically. 
A crucial number is the power spectrum of the perturbations,
which in the long wavelength limit (it is the most important 
for cosmology) is of the type $\,P^2_n \propto n^k$. 
This distribution tells how the amplitude of the perturbations
depends on $n$. The coefficient
$k$ is called spectral indice of the perturbation.

In the standard DeSitter case \cite{gri} one meets exact result
$k = 0$. The Cobe, Boomerang and Maxima observational 
programs favor the flat spectrum too, with the bounds
$\,\,-0.15 < k < 0.16$ (see, for example, \cite{exp}).
For our model, the numerical procedure gives \cite{wave}
$\,k \simeq - 0.01 \,$
which is qualitatively in agreement with a flat spectrum.

Hence, our model
predicts a spectral indice different, but not very far
from the one predicted by the traditional inflationary scenario.
Besides, the prediction has quite a good agreement with the
observational results. 

As a conclusion, we consider the stable anomaly-induced model
\cite{anju,wave} as the most natural one, because it does not 
require any fine-tuning, nor for inflation neither for the 
spectrum of metric perturbations. The only serious problem 
is the grace-exit, which can be hopefully solved in the 
framework of effective approach.


\end{document}